\documentclass[12pt]{article}
\usepackage{amsxtra,amssymb,amsthm,amsmath,latexsym}

\theoremstyle{plain}

\newcommand{\refS}[1]{Section~\ref{S:#1}}

\def\bysame{\rule{.5in}{.005in},\ }

\def\ve{{\varepsilon}}
\def\R{{\mathbb R}}

\def\calA{{\mathcal A}}\def\calS{{\mathcal S}}\def\calU{{\mathcal U}}

\def\tildeb{\widetilde\beta}\def\tildeg{\widetilde\gamma}
\def\oH{{\overset{\circ}{H}}}\def\oH1{{\overset{\circ}{H}\kern-.02in{}^1}}

\def\bee{\begin{equation*}} \def\eee{\end{equation*}}
\def\be{\begin{equation}}\def\ee{\end{equation}}

\begin{document}%\begin{titlepage}
\title{Many-body wave scattering by small bodies}
\author{A.G. Ramm\\
 Mathematics Department, Kansas State University, \\
 Manhattan, KS 66506-2602, USA\\
ramm@math.ksu.edu,\\ fax 785-532-0546, tel. 785-532-0580\\
http://www.math.ksu.edu/\,$\widetilde{\ }$\,ramm}
\date{}\maketitle\thispagestyle{empty}
\begin{abstract}
\footnote{PACS: 0340K, 4110H, 4320}
\footnote{key words: many-body scattering, acoustic wave scattering,
electromagnetic wave scattering, numerical analysis}

Scattering problem by several bodies, small in comparison with the wavelength, 
is reduced to linear algebraic systems of equations, in contrast to the 
usual reduction
to some integral equations.
\end{abstract}%\end{titlepage}

\section{Introduction}\label{S:1} 

Acoustic or electromagnetic (EM) wave scattering by one or several bodies 
is
usually studied by reducing the problem to solving some integral
equations. In this paper we show that if the bodies are small in
comparison with the wavelength, then the scattering problem can be reduced
to solving linear algebraic systems with matrices whose elements have
physical meaning. These elements are electrical capacitances or
elements of electric and magnetic polarizability tensors. The author
has derived analytical explicit formulas  allowing
one to calculate these quantities for bodies of arbitrary shapes with 
arbitrary desired accuracy (see  \cite{R476}).

 We derive these linear algebraic systems and give formulas for the
elements of the matrices of these systems. There is a large literature on
wave scattering by small bodies, see \cite{R476} and references therein.
The theory was originated by Lord Rayleigh \cite{Ray}, who understood
that the main term in the scattered field is the dipole radiation if the
body is small. Rayleigh did not give formulas for calculating the induced
dipole moments for small bodies of arbitrary shapes.
The dipole moments are uniquely defined by the polarizability tensors.
Therefore, the formulas,  
derived by the author (see \cite{R476}), allow one to calculate
the dipole radiation for acoustic and EM wave scattering by small bodies 
of arbitrary 
shapes.

\section{Acoustic wave scattering by small bodies}\label{S:2}

Let us start with acoustic wave scattering. Consider the problem
\be\label{e1}
 (\Delta+k^2)u=0\hbox{\quad in\quad}\R^3\setminus(\bigcup^m_{m=1}D_m) \ee 
\be\label{e2}
 u\mid_{S_m}=0, \qquad 1\leq m\leq M, \quad S_m:=\partial D_m\ee
\be\label{e3}  u=u_0+v,\ee
\be\label{e4}
  \frac{\partial v}{\partial r}-ikv=o\left(\frac1r\right),
  \qquad r:=|x|\to\infty,\ee

where $\Delta$ is the Laplacean, $u_0$ is an incident field which solves 
equation \eqref{e1}. Often, 
$u_0=e^{ik\alpha\cdot x}$, where $\alpha\in S^2$ is a given vector and 
$S^2$ is 
the unit sphere.

Let us look for the solution of the form
\be\label{e5}
 u=u_0+\sum^M_{m=1}\int_{S_m} g(x,s)\sigma_m(s)ds,
 \qquad g(x,y):=\frac{e^{ik|x-y|}}{4\pi|x-y|}_, \ee
where $\sigma_m$, $1\leq m \leq M$, are to be chosen so that the boundary 
conditions 
\eqref{e2} hold.
The function \eqref{e5} satisfies \eqref{e1} and \eqref{e3}- \eqref{e4} 
for any
 $\sigma_m \in L^2(S_m)$. 
The scattering amplitude is:
\be\label{e6}
 \calA (\alpha',\alpha)=\lim_{|x|\to\infty,\, \frac{x}{|x|}=\alpha'} |x| 
e^{-ik|x|}\,
 v=\sum^M_{m=1}\frac{1}{4\pi}\int_{S_m} e^{-ik\alpha'\cdot s}\sigma_m 
ds,
 \quad \alpha':=\frac{x}{|x|}. \ee
Let \[a:=\max_{1\leq m \leq M}diam\,D_m,\] and \[d:=\min_{m\neq j} 
dist(D_m, D_j).\] We assume
\be\label{e7} ka\ll1, \qquad a\ll d. \ee
Then \[e^{-ik\alpha'\cdot(s-x_m)}\approx 1 \hbox{\quad if \quad} x_m\in 
D,\] so
\be\label{e8}
 \calA(\alpha',\alpha)= \sum^m_{m=1} \frac{e^{-ik\alpha'\cdot x_m}}{4\pi} 
 \int_{S_m}\sigma_m ds:=\sum^M_{m=1} \frac{Q_m}{4\pi}
  \,e^{-ik\alpha'\cdot x_m},\quad Q_m:=\int_{S_m} \sigma_m ds, \ee
where $x_m\in D_m$ and $\alpha'$ is defined in \eqref{e6}. Since $D_m$ is 
small, it does not matter which 
point 
$x_m$ one takes in $D_m$.
The $Q_m$ plays the role of the total charge on the surface $S_m$.

If $\min_m|x-x_m|\gg a$ and $x_m\in D_m$, then
\be\label{e9}
 u(x)=u_0(x)+\sum^M_{m=1} \frac{e^{ik|x-x_m|}}{4\pi|x-x_m|}\,
 Q_m \left[1+O\left( ka+\frac{a}{d}\right)\right].\ee
Let us derive a formula for $Q_m$. Using the boundary condition  
\eqref{e2}, one gets:
\be\label{e10}
0=u_0(s_m)+\sum_{j\not=m}g(s_m,x_j)Q_j+\int_{S_m}g(s_m,s)\sigma_m(s)ds,\ee
where $s_m\in S_m$.

Since $ka\ll 1$, one has \[g(s_m,s)=g_0(s_m,s)+O(ka),\] where
\[g_0(s,t):=\frac{1}{4\pi|x-t|}.\]
Therefore equation \eqref{e10} is the equation for the electrostatic 
charge 
distribution $\sigma_m$ on the surface $S_m$ of a perfect conductor $D_m$, 
charged to the potential 
\[u_m:=-u_0(s_m)-\sum_{j\not= m} g(s_m,x_j)Q_j.\]
The total charge on $S_m$ is:
\[Q_m=C_m \calU_m,\]
where $C_m$ is the electrical capacitance of the conductor with the 
shape $D_m$. The total charge is defined as:
\[Q_m:=\int_{S_m}\sigma_m ds.\] 
Therefore, one gets:
\be\label{e11}
 Q_m=C_m\left(-u_0(s_m)-\sum_{j\not= m}g(s_m,x_j)Q_j\right),
 \qquad 1\leq m\leq M, \ee
where $C_m$ is the electrical capacitance of the perfect conductor with 
the boundary $S_m$.

Linear algebraic system \eqref{e11} allows one to find $Q_j$, $1\leq j\leq 
M$. If
\be\label{e12}
 \max_{1\leq m\leq M} \sum_{j\not=m}\ \frac{C_m}{4\pi|s_m-x_j|}<1, \ee
then the matrix of the system \eqref{e11} has diagonally dominant elements 
and, consequently, can be solved by iterations.

The approximate solution to the many-body scattering problem 
\eqref{e1}--\eqref{e4} is given by formula \eqref{e9}, where $Q_m$ are 
determined from linear algebraic system \eqref{e11}. 

Let us give a formula from \cite{R476}
for the capacitance of a perfect conductor $D$ with the boundary $S$.  
Denote the area of $S$ by   
$|S|$. We assume that the conductor is placed in the medium with the 
dielectric permittivity 
$\ve_0=1$. In this case the approximate formula for the capacitance
is (see 
\cite{R476}, p. 26):
\be\label{e13}
      C^{(n)} = 4 \pi |S|^2
      \left\{ 
      \left( \frac{-1}{2 \pi}\right)^n \int_S\int_S \frac{dsdt}{r_{st}}
    \underbrace{\int_S\dots\int_S}_{n \hbox{\begin{tiny}\ times\end{tiny}}}  
      \psi(t,t_1) \dots \psi (t_{n-1}, t_n) dt_1 \dots dt_n
      \right\}^{-1}_,
      \ee
\[  
\begin{aligned}
   C^{(0)}
   & = \frac{4 \pi |S|^2}{J} \leq C,
   \qquad J:= \int_S \int_S\frac{dsdt}{r_{st}},
   \qquad r_{st}:=|s-t|,\\
   &\psi(t,s)= \frac{\partial}{\partial N_t}\frac 1 {r_{st}}, 
\end{aligned}
\]  
and the error estimate of formula \eqref{e13} is:
\be\label{e14}
 \qquad |C^{(n)}-C|=O(q^n),
   \qquad 0<q<1,
   \ee 
where $q$ depends on the geometry of $S$, and $n=1,2,3.....$ is the 
approximation order. 

If the boundary condition 
\be\label{e15}
 u_N=\zeta u \hbox{\quad on\quad} S_m \ee
is imposed in place of the Dirichlet condition \eqref{e2},
and $\zeta$ is the impedance, then $C_m$ in 
\eqref{e11} is replaced by
\be\label{e16}
 C_{m\zeta}:=\frac{C_m}{1+C_m(\zeta|S|)^{-1}}, \ee
see \cite{R476}, p. 97.

If 
\be\label{e17}
 u_N\mid_{S_m}=0, \qquad 1\leq m\leq M, \ee
then the formula for the solution to problem \eqref{e1}, \eqref{e17}, 
\eqref{e3}, \eqref{e4}, is
\be\label{e18}
 u(x)=u_0(x)+\sum^M_{m=1} q(x,x_m) V_m
 \left[\Delta u(x_m)+\sum^M_{p,q=1} \beta_{pq,m} 
  ik\frac{\partial u(x_m)}{\partial x_{m,q}}\ 
\frac{(x-x_m),_{p}}{|x-x_m|}, 
 \right]\ee
where $(x-x_m),_{p}$ is the $p\hbox{-th}$ 
coordinate of the vector $x-x_m$,
$\frac{\partial}{\partial x_{m,q}}$ is the derivative with respect to the
$q\hbox{-th}$ coordinate of $x$ calculated at the point $x_m$, 
and $\beta_{pq,m}$ is the magnetic polarizability tensor of $D_m$, defined 
by the formula  ([1], p.98):
\bee V_m \beta_{pq,m}=\int_S s_p\sigma(s)ds, \eee
where $ V_m$ is the volume of $D_m$, the function $\sigma$ solves the 
equation \[\sigma=A\sigma-2N_q,\]
 $N$ is the 
exterior unit normal to $S_m$, and 
\[A\sigma=\int_{S_m}\frac{\partial}{\partial N}
 \frac{1}{2\pi r_{st}}\sigma(t)dt, \qquad r_{st}=|s-t|.\]

The formulas for the tensor $\beta_{pq,m}$, analogous to the formulas 
\eqref{e13}-\eqref{e14} for the capacitance, are derived in 
\cite[p.55, formula (5.15)]{R476}. The unknown quantities $\Delta u(x_m)$ 
and $\frac{\partial u(x_m)}{\partial x_q}$, $1\leq m\leq M$, 
$1\leq q\leq 3$, in \eqref{e18} can be found from the following linear 
algebraic system, analogous to \eqref{e11}:
\be\label{e19}
 \Delta u(x_m)=\Delta u_0(x_m)- k^2 \sum^M_{j\not= m, j=1}g(x_m,x_j)V_j
% \left
[ \Delta u(x_j)+\sum^3_{p,q=1} \beta_{pq,j} ik 
 \frac{\partial u(x_j)}{\partial x_{j,q}}
 \frac{(x_m-x_j),_{p}}{|x_m-x_j|} 
%\right
] 
\ee
\be\label{e20}
 \frac{\partial u(x_m)}{\partial x_{m,q}}
 =\frac{\partial u_0(x_m)}{\partial x_{m,q}}
 +\sum^M_{j\not=m,j=1} \frac{\partial g(x_m,x_j)}{\partial x_{m,q}}
 V_j 
%\left
[ \Delta u(x_j)+\sum^3_{p,q=1} \beta_{pq,j}
 ik \frac{\partial u(x_j)}{\partial x_{j,q}}
 \frac{(x_m-x_j),_{p}}{|x_m-x_j|} 
%\right
]. \ee
In \eqref{e19} we have used the equation \[\Delta g(x,y)=-k^2g(x,y),\] 
which holds if $x\not=y$.

From the linear algebraic system \eqref{e19}--\eqref{e20} one finds the 
unknowns $\Delta u(x_m)$ and $\frac{\partial u(x_m)}{\partial x_{m,q}}$, 
$1\leq m\leq M$, $1\leq q\leq 3$.

If conditions \eqref{e7} hold, then system \eqref{e19}--\eqref{e20}
has a unique solution which can be obtained by iterations.

This completes the description of our method for solving many-body 
scattering problem for small bodies and acoustic (scalar) waves.

\section{Electromagnetic wave scattering by small bodies}\label{S:3}

In the problem of electromagnetic (EM) wave scattering by many small 
bodies we assume
\be\label{e21}  a\ll \lambda \ll d. \ee
This assumption is more restrictive than \eqref{e7}. The reason is: in EM 
theory the fields are obtained by an application of first order 
differential operators, for instance $\nabla \times $, to 
potentials, such 
as the vector potential.
  Applying this operator and calculating the field in the far zone one 
neglects the term 
$|\frac{1}{x-x_m}|$ compared with the term $k$. This means that
the following inequality is assumed: 
\[\frac1d\ll\frac1\lambda,\] or 
\[d\gg\lambda.\] 
In the acoustic wave theory the potential itself $\int_S g(x,s)\sigma
ds$ has physical meaning, it is the acoustic pressure, and this  
pressure is studied. Therefore, the
condition $d\gg\lambda$ does not appear. 

Condition \eqref{e7} allows
one to have many small particles on the distance of order $\lambda$,
while condition \eqref{e21}, namely the inequality $d\gg\lambda$, does not 
allow this.
Recall that $d$ is the minimal distance between two neighboring
particles. The formula for the scattering amplitude, analogous to 
\eqref{e8}, for EM wave scattering
by small bodies is (see \cite{R450}): 
\be\label{e22} 
   A(\theta',\theta)=\frac{1}{4\pi}
   \sum^M_{m=1}\calS_m \calU_me^{-ik\theta'\cdot x_m}.\ee
Here $\calU=\binom{E}{H}$ is a 6-component vector,  $\calS_m$ is 
a 6x6 matrix, the scattering matrix,
 $\ve_0$ and $\mu_0$ are
dielectrical and magnetic parameters of the medium, in which the body 
$D_m$ is placed, and
 $\theta,\theta'$ are the unit vectors 
in the direction of the 
incident and scattered waves, respectively. These vectors were denoted 
$\alpha$ and 
$\alpha'$ in \refS{2}. We have changed the notations because in EM theory 
$\alpha$ denotes the polarizability tensor. 

The formula for $\calS$ is (cf. [2])
\be\label{e23}
 \calS_m\binom{E}{H} =\frac{k^2V_m}{4\pi}
  \begin{pmatrix}
   \alpha E-\theta'(\theta',\alpha E)
   &-\frac{\mu^{3/2}_0}{\ve^{1/2}_0}[\theta',\tildeb H]\\
   \left(\frac{\ve_0}{\mu_0}\right)^{\frac12}[\theta',\alpha E]
   &\quad\mu_0(\tildeb-\theta'(\theta',\tildeb H))
   \end{pmatrix}_.\ee
Here $V_m$ is the volume of $D_m$, $\alpha$ is the electric polarizability 
tensor of $D_m$, $\tildeb$ is the magnetic polarizability tensor  of 
$D_m$. In 
\cite[pp.~ 54--55]{R476} the author derives analytical formulas for 
calculation of the polarizability tensors $\alpha$ and $\beta$,
\be\label{e24}
 \tildeb:=\alpha(\tildeg)+\beta, \quad \beta:=\alpha_{ij}(-1),
 \quad\tildeg:=\frac{\mu-\mu_0}{\mu+\mu_0}, 
 \quad \gamma:=\frac{\ve-\ve_0}{\ve+\ve_0}. \ee
Tensors $\beta$ and $\tildeb$ are expressed through
the polarizability tensor $\alpha=\alpha(\gamma)$. One has 
$\gamma=-1$ if $\ve=0$. Here 
$[\cdot,\cdot]$ is the vector product, $(\cdot,\cdot)$ is the 
scalar product.

The analytic formula from \cite{R476}, p. 54, formula (5.9), for the 
tensor 
$\alpha=\alpha_{ij}(\gamma)$, $1\leq i$, $j\leq 3$, that we referred to 
above, is analogous to formulas \eqref{e13}--\eqref{e14} for the 
electrical 
capacitance. The incident direction $\theta$ enters via the vectors $E$ 
and $H$, which depend on $\theta$. These vectors are calculated
in formula \eqref{e23} at the 
point $x_m$. The values of these vectors are 
determined from a linear algebraic system of equations. This system is 
derived similarly to the derivation of the systems \eqref{e11} and 
\eqref{e19}--\eqref{e20}. We do not write down this system since it 
would 
take much space, but the ideas are the same as the ones used in the 
derivations of \eqref{e11} and \eqref{e19}--\eqref{e20}.

\section{Conclusions}

In this paper it is shown how to reduce rigorously the many-body 
scattering 
problem to linear algebraic system in the case when the bodies are small 
in comparison with the wavelength. The theory is constructed for acoustic 
and EM wave scattering. The basic physical assumptions are \eqref{e7} for 
acoustic scattering, and \eqref{e21} for EM scattering.

\end{document}